\documentclass[doublecol]{epl2}

\newcommand{\ba}[1]{\bm{a}^{(#1)}}
\newcommand{\bH}[1]{{\cal H}^{(#1)}}

\newcommand{\bu}{\bm{u}}

\newcommand{\bdelta}{\bm{\delta}}

\newcommand{\bxi}{\bm{\xi}}

\newcommand{\dd}[2]{\frac{d #1}{d #2}}
\newcommand{\pp}[2]{\frac{\partial #1}{\partial #2}}

\title{Galilean invariance of lattice Boltzmann models}
\author{X. B. Nie \and X. Shan \and H. Chen}
\institute{Exa Corp., 3 Burlington Woods Drive, Burlington, MA 01803, USA}


\abstract{It is well-known that the original lattice Boltzmann (LB)
equation deviates from the Navier-Stokes equations due to an
unphysical velocity dependent viscosity.  This unphysical dependency
violates the Galilean invariance and limits the validation domain of
the LB method to near incompressible flows.  As previously shown,
recovery of correct transport phenomena in kinetic equations depends
on the higher hydrodynamic moments.  In this Letter, we give specific
criteria for recovery of various transport coefficients.  The Galilean
invariance of a general class of LB models is demonstrated via
numerical experiments.}

\begin{document}

\maketitle

\section{Introduction}

The idea behind the discrete-velocity kinetic method is that the full
details of the single-particle distribution function in the
three-dimensional microscopic velocity space is not entirely necessary
in describing the thermo-hydrodynamics of a fluid.  Instead, much
simpler kinetic systems in which the constituent particles can only
have velocities from a small discrete set can be devised to exhibit
the same macroscopic behavior.  This idea can be traced back to the
works of Joule.  It was later exploited by
Broadwell~\cite{Broadwell64a,Broadwell64b} and more recently has lead
to the development of the lattice gas cellular automaton (LGA) and
lattice Boltzmann (LB)
models~\cite{Doolen87,Succi01,McNamara88,Higuera89b}.  The key
question concerning the validity of this class of models is how
closely the macroscopic thermo-hydrodynamics of the continuum kinetic
system can be reproduced by the much simplified kinetic systems.

It is well known that the original LGA~\cite{Frisch86,Wolfram86} has a
non-Galilean-invariant hydrodynamics due to the artifacts in the
advection term and the equation of state.  The lattice BGK (LBGK)
method~\cite{Qian92a,Chen92a} eliminated these artifacts by choosing a
proper equilibrium distribution function in the single-relaxation-time
(BGK) collision term.  The Navier-Stokes hydrodynamics at the
small-Mach-number limit is reproduced.  Despite its phenomenal success
in modeling near-incompressible athermal
flows~\cite{Shan93,Nie02,Chen03,Sbragaglia07}, the LBGK method is still not fully
Galilean invariant due to a ``cubic'' velocity dependency of the
viscosity~\cite{Qian93}, which is one of the factors that limit the
valid domain of LB method to near-incompressible flows.  Several
attempts~\cite{Chen94,Qian98,Hazi06} have been made to overcome this
short-coming using more velocities and higher order terms in the
Mach-number expansion of the equilibrium distribution.  The success is
however limited for that not only the process of obtaining the
high-order correction is tedious and model-dependent, but also the
Galilean invariance is in some cases only partially restored.

Recently, the LB method was re-formulated as a Hermite series solution
to the continuum BGK equation~\cite{Shan98,Shan06}, in a way similar
to the Grad 13-moment system.  Under this formulation, the
Chapman-Enskog~\cite{Chapman70} procedure can be applied to the LBGK
system in the Hermite space, revealing that the convergence of LB to
continuum BGK relies on the few leading moments of the equilibrium
distribution function.  Provided that these moments agree with those
of the Maxwell distribution, the macroscopic hydrodynamics of the LBGK
equation is the same as that of the continuum BGK equation, which is
known to be fully Galilean invariant.  Once scrutinized in this
framework, the commonly known LB models are insufficient in meeting
the above conditions of retaining hydrodynamic moments, and therefore
introduce errors such as the well-known cubic velocity dependency of
the viscosity.  This analysis is general enough so that it can be
applied to the energy equation, the diffusion equation in a
multi-component system, and higher hydrodynamic approximations beyond
the Navier-Stokes level to give convergence criteria of other
transport coefficients at the Navier-Stokes level and beyond.

In this Letter, we define systematically the necessary and sufficient
conditions for the LBGK hydrodynamics to converge to that of the
continuum BGK.  Essentially, these conditions are that sufficient
hydrodynamic moments must be retained in the equilibrium distribution
of the LBGK system and accurately represented by the discrete
velocities.  We also present numerical simulation results that
validate these conditions.  In the next section, we first
give a more elaborated derivation of the Chapman-Enskog calculation in
Hermite space, and then point out the convergence condition as the
direct consequence.  These convergence
conditions are then verified numerically by directly measuring the
transport coefficients in LB simulations.  Further discussions are
offered in the last section.

\section{Convergence conditions}

\label{sec:hermite}

In a previous publication~\cite{Shan06}, the asymptotic convergence of
LB to the continuum BGK was briefly shown using Chapman-Enskog
approximation procedure in Hermite space.  Same as in continuum
kinetic theory, the macroscopic equations of LB are essentially
determined by the leading tensorial hydrodynamic moments.  With
discrete velocities, these moments are expressed as weighted sum of
tensors constructed from discrete velocities, which are not always
isotropic as their continuum counterparts are.  The insufficient
isotropy of certain moments has long been recognized as one of the
reasons responsible for the deviations from the continuum
hydrodynamics.  By expanding the distribution function in Hermite
polynomials, this tedious calculation can be tremendously simplified
and extended to higher orders.

We first note that the hydrodynamic equations are given by the
conservations of mass, momentum and energy.  Omitting external forces,
they are:
\begin{eqnarray}
  \label{eq:mass}
  &&\dd\rho t + \rho\nabla\cdot\bu = 0\\
  \label{eq:momentum}
  &&\rho\dd\bu t + \nabla\cdot\bm{P} = 0\\
  \label{eq:energy}
  &&\rho\dd\epsilon t + \bm{P}:\nabla\bu + \frac 12\nabla\cdot\bm{S} = 0,
\end{eqnarray}
where $d/dt$ represents substantial derivative and the
density $\rho$, fluid velocity $\bu$, internal energy density per mass
$\epsilon$, pressure tensor $\bm{P}$ and heat flux $\bm{S}$ are all
velocity moments of $f$:
\begin{eqnarray}
  \label{eq:moments}
  \rho &=& \int fd\bxi\\
  \rho\bu &=& \int f\bxi d\bxi\\
  \rho\epsilon &=& \frac 12\int f|\bxi-\bu|^2 d\bxi\\
  \bm{P} &=& \int f(\bxi-\bu)(\bxi-\bu)d\bxi\\
  \bm{S} &=& \int f|\bxi-\bu|^2(\bxi-\bu)d\bxi,
\end{eqnarray}
where $f$ is the single-particle distribution function and $\bxi$ the
microscopic velocity.  Clearly, two distribution functions with the
same leading velocity moments will yield the same hydrodynamic
equations.

For closing Eqs.~(\ref{eq:mass})--(\ref{eq:energy}), the task of
obtaining the approximated distribution function using the few
hydrodynamic moments was in the center of the development of the
kinetic theory in the last century.  In the Chapman-Enskog
calculation, a successive sequence of approximated distribution
functions are introduced in the order of Knudsen
number~\cite{Chapman70}.  For the following Boltzmann equation with
the BGK collision model~\cite{Bhatnagar54}:
\begin{equation}
  \label{eq:bgk}
  \pp ft + \bxi\cdot\nabla f = -\frac 1\tau\left[f-f^{(0)}\right],
\end{equation}
where $\tau$ is the relaxation time, the $i$-th approximation of the
distribution is given by the recursive relation:
\begin{equation}
  \label{eq:CE}
  f^{(i)} = -\tau\left(\pp{}t+\bxi\cdot\nabla\right)f^{(i-1)},\quad
  i = 1,2, \cdots.
\end{equation}
with $f^{(0)}$ is the Maxwell-Boltzmann distribution.  On substituting
$f^{(i)}$ into Eqs.~(\ref{eq:mass})--(\ref{eq:energy}) for $i=0$, $1$
and $2$ and using results of the previous approximation, the
hydrodynamic equations at the Euler, the Navier-Stokes, and the
Burnett levels are obtained~\cite{Huang87}.  This sequence of
approximation quickly becomes formidably complex beyond the first
two.

Eq.~(\ref{eq:CE}) implies a simple recursive relation among the
moments of the distribution functions at various approximation level.
We note that the velocity moments of the distribution function are
essentially its Hermite coefficients~\cite{Grad49}.  Let $\ba{n}_i$ be
the $n$-th Hermite coefficient of the distribution $f^{(i)}$.  As
previously shown, on substituting the corresponding Hermite expansions
into Eq.~(\ref{eq:CE}), the Hermite coefficients of the first
approximation can be explicitly expressed using the Hermite
coefficients of the zero-th approximation, namely:
\begin{equation}
  \label{eq:lb}
  \ba{n}_1 = -\tau\left[\pp{\ba{n}_0}t + \nabla\ba{n-1}_0 +
    \nabla\cdot\ba{n+1}_0\right].
\end{equation}
>From the above equation, an important conclusion that can be
immediately drawn about the BGK equation (\ref{eq:bgk}) is that the
leading $n$ moments in the first approximation, which gives the
Navier-Stokes hydrodynamics, is given by the leading $n+1$ moments of
$f^{(0)}$.  The form of the hydrodynamic equations is completely
determined by the few leading moments of $f^{(0)}$.  For instance, at
the Navier-Stokes level ($i=1$), $\bm{P}$ is determined by $\ba{2}_1$,
which is in turn determined by the expansion coefficients of $f^{(0)}$
up to $\ba{3}_0$.  Therefore, the momentum equation is correct as long
as $f^{(0)}$ agrees with the Maxwellian in the first three terms of
their Hermite expansions.

For a finite sum of Hermite series, the leading moments are completely
determined by the function values at a finite set of abscissas.  This
property, together with the fact that the hydrodynamic equations are
determined by the finite number of leading Hermite terms of $f^{(0)}$,
allows a discrete-velocity kinetic system to have the same macroscopic
hydrodynamics as the continuum system when their leading moments are
the same.  Consider a finite sum of Hermite series as $f^{(0)}$:
\begin{equation}
\label{feq}
  f^{(0)} = \omega(\bxi)\sum_{n=0}^N\frac 1{n!}\ba{n}_0\bH{n}(\bxi).
\end{equation}
Let $M$ be the highest order of the moments that determines the
hydrodynamic equations of interest, {\it e.g.}, $M=3$ if the momentum
equation at the Navier-Stokes level is of concern.  The leading $M$
($\leq N$) moments are in the form of
\begin{equation}
  \int\omega(\bxi)\bm{p}(\bxi)d\bxi,
\end{equation}
where $\bm{p}(\bxi)$ is a polynomial of an order $\leq M+N$.  This
integral can be exactly evaluated using the values of the integrand on
a finite set of velocities, $\{\bxi_i: i=1, \cdots, d\}$, as:
\begin{equation}
  \label{eq:quad}
  \int\omega(\bxi)\bm{p}(\bxi)d\bxi=\sum_{i=1}^dw_i\bm{p}(\bxi_i),
\end{equation}
if and only if $\bxi_i$ are the abscissas of a Gauss-Hermite
quadrature of a degree of precision $Q \geq M+N$, and $w_i$ the
corresponding wights.  It can now be concluded that the
sufficient conditions for the LBGK system to have the correct
hydrodynamic equations are:
\begin{enumerate}
  \item The equilibrium distribution retains the necessary
  moments.
  \item The discrete velocity set allows the moments to be exactly
  evaluated using finite function values.
\end{enumerate}
Quantitatively, these two conditions are:
\begin{equation}
   \label{eq:mn}
  N\geq M \quad\mbox{and}\quad M+N \leq Q.
\end{equation}
Since $Q$ is finite for any finite set of velocities and $N\leq Q$,
$N$ must be finite, {\it i.e.}, the equilibrium distribution in LBGK
is the sum of a {\em finite} Hermite series of an order which is
smaller than the degree of the quadrature.  The difference between $Q$
and $N$ is the maximum order of the moments whose dynamics will be the
same as that in the continuum system.

The requirements for the momentum equation to be fully Galilean
invariant at Navier-Stokes level is immediately clear.  Since $M=3$ in
this case, we have $N\geq 3$ and $Q\geq 6$.  Namely $f^{(0)}$ must
retain all Hermite terms up to the third order and the discrete
velocity set must form a 6th-order accurate Hermite quadrature.
Neither of the original LBGK models~\cite{Qian92a,Chen92a} satisfies
this condition as only the second order terms are retained and the
velocity set is only 5-th order in both cases ($Q=5$).  As the
velocity set is sufficient to support the second-order terms, the
error introduced in both models are due to the last term on the
right-hand-side of Eq.~(\ref{eq:lb}), which manifests as the ``cubic''
velocity-dependence of the viscosity~\cite{Qian93,Shan06}.

The analysis above can be applied to the case of energy equation where
full recovery of the Navier-Stokes energy equation requires $N\geq
M\geq 4$, and $Q\geq 8$.  Using a third-order expansion ($N=3$), or a
velocity set that only supports third-order moments ($6\leq Q< 8$)
could yield otherwise correct thermal LBGK models with a
velocity-dependent thermal diffusivity.

It is worth pointing out that when conditions (\ref{eq:mn})
are satisfied, the equilibrium function of Eq.~(\ref{feq}) results in
the exact hydrodynamics equations regardless of Mach number.
Hydrodynamic moments up to a given order can be shown to have the same
dynamics of their continuum counterparts.  These conditions are also
necessary if a finite order polynomial is used as the the equilibrium
distribution.
%When an arbitrary function containing Hermite terms of all orders is used as the equilibrium distribution, no link can be established between the integrals in continuum velocity space and the summations in the discrete velocity space.
Nevertheless, it was shown
previously~\cite{Ansumali03,Chikatamarla06} that discrete velocity
models can be constructed with exact conservation laws and correct
higher moments in the small Mach number limit.

We note that for a given set of velocities and weights, ${\cal
Q}\equiv \{(\bxi_i, w_i):i=1, \cdots, d\}$, the isotropy of the
following tensors play a critical role in the analysis of LGA and
LBGK~\cite{Wolfram86}:
\begin{equation}
  \bm{E}^{(n)} = \sum_{i=1}^dw_i
  \underbrace{\bxi_i\cdots\bxi_i}_{\mbox{$n$ times}}.
\end{equation}
Here, we point out that ${\cal Q}$ corresponding to a $Q$-th order
Gauss-Hermite quadrature is equivalent to the condition that:
\begin{equation}
  \label{eq:int}
  \bm{E}^{(n)} = \left\{\begin{array}{ll}
    0 & \mbox{$n$ odd}\\
    \bdelta^{(n)} &  \mbox{$n$ even}
  \end{array}\right.,\quad\forall n \leq Q,
\end{equation}
where $\bdelta^{(n)}$ is the rank-$n$ isotropic tensor.  Noticing that
the above is always true in continuum, {\it i.e.}, after defining
$\bm{p}_n(\bxi)\equiv\underbrace{\bxi\cdots\bxi}_{\mbox{$n$ times}}$,
we have:
\begin{equation}
  \int\omega(\bxi)\bm{p}_n(\bxi)d\bxi = \left\{\begin{array}{ll}
    0 & \mbox{$n$ odd}\\
    \bdelta^{(n)} & \mbox{$n$ even}
  \end{array}\right.,
\end{equation}
the forward equivalence is trivial since $\forall n\leq Q$,
$\bm{p}_n(\bxi)$ is a polynomial of a degree $\leq Q$.  Because of
Eq.~(\ref{eq:quad}),
\begin{equation}
  \bm{E}^{(n)} = \int\omega(\bxi)\bm{p}_n(\bxi)d\bxi =
  \left\{\begin{array}{ll}
  0 & \mbox{$n$ odd}\\
  \bdelta^{(n)} & \mbox{$n$ even}
  \end{array}\right..
\end{equation}
The backward equivalence is also straightforward.  As
$\{\bm{p}_n(\bxi): n\leq Q\}$ is a linearly independent set and forms
a complete basis of the Hilbert space of all polynomials of a degree
$\leq Q$, any such polynomial can be written as a linear combination
of $\bm{p}_n(\bxi)$.  By direct integration, it can be shown that
Eq.~(\ref{eq:quad}) is true as long as Eq.~(\ref{eq:int}) is true.
Condition (\ref{eq:int}) is also proved by a slightly different
procedure elsewhere~\cite{Chen07}.

\section{Numerical simulations}

\label{sec:numerics}

To numerically demonstrate the theory outlined in the previous section,
here we present simulation results of some simple benchmarking
problems using LBGK models with equilibrium distributions and velocity
sets of different orders.  All velocity sets are in three-dimensions,
and they are the well-known D3Q19 model ($Q=5$)~\cite{Qian92a}, the
seventh-order ($Q=7$) 39-velocity quadrature $E^{19}_{3,7}$ of
Ref.~\cite{Shan06} ($Q=9$) and a ninth-order 121-velocity quadrature
$E^{121}_{3,9}$.  The two-dimensional projection of the last model was
used in a previous study~\cite{Shan07}.  Here given in
Table~\ref{tb:q3d} are the full detail of the three-dimensional
version.  All simulations are performed on a $1\times1\times 100$ grid
in Cartesian coordinates $\{x,y,z\}$ with periodic boundary condition
used in all directions.

\begin{table}[ht]
  \caption{The abscissas and weights of a ninth-order accurate
    Gauss-Hermite quadrature formula in three-dimensions.  Here $p$
    is the number of abscissas in the symmetry class.  The subscript
    $_{FS}$ denotes a fully symmetric set of points.}
  \begin{tabular}{@{}rrc@{}}
	\hline
    $\bxi_a$                 &$p$ & $w_a$    \\
	\hline
    $(0, 0, 0)$              &  1 & 0.03059162202948600642469 \\
    $(r, 0, 0)_{FS}$         &  6 & 0.09851595103726339186467 \\
    $(\pm r,\pm r, \pm r)$   &  8 & 0.02752500532563812386479 \\
    $(r, 2r, 0)_{FS}$        & 24 & 0.00611102336683342432241 \\
    $(2r,2r, 0)_{FS}$        & 12 & 0.00042818359368108406618 \\
    $(3r, 0, 0)_{FS}$        &  6 & 0.00032474752708807381296 \\
    $(2r,3r, 0)_{FS}$        & 24 & 0.00001431862411548029405 \\
    $(\pm 2r,\pm 2r,\pm 2r)$ &  8 & 0.00018102175157637423759 \\
    $(r, 3r, 0)_{FS}$        & 24 & 0.00010683400245939109491 \\
    $(\pm 3r,\pm 3r,\pm 3r)$ &  8 & 0.00000069287508963860285 \\
	&& $r = 1.19697977039307435897239$\\
	\hline
  \end{tabular}
  \label{tb:q3d}
\end{table}

To measure the velocity-dependence of the viscosity, we simulated the
one-dimensional shear wave problem where both density and temperature $T_0$
are constants initially.  The initial velocity is given by:
\begin{equation}
  \bu = \bu_0 + a_0\bm{e}_x\sin(2\pi z/L_z),
\end{equation}
where $\bu_0$ is a homogeneous constant translational velocity field,
$a_0$ the small initial amplitude of the shear, $\bm{e}_x$ the unit
vector in the $x$ direction, and $L_z = 100$ the periodicity in $z$
direction.  The initial distribution function $f$ is assigned with its equilibrium
value based on the initial density, temperature and velocity. 
When the Navier-Stokes equations are fully satisfied, the
amplitude of the shear wave shall decay exponentially independent of
$\bu_0$ with a decay rate proportional to the viscosity.  The
viscosity is measured through the decay rate of the amplitude.

\begin{figure}
  \onefigure[viewport=50 50 554 928,width=240pt]{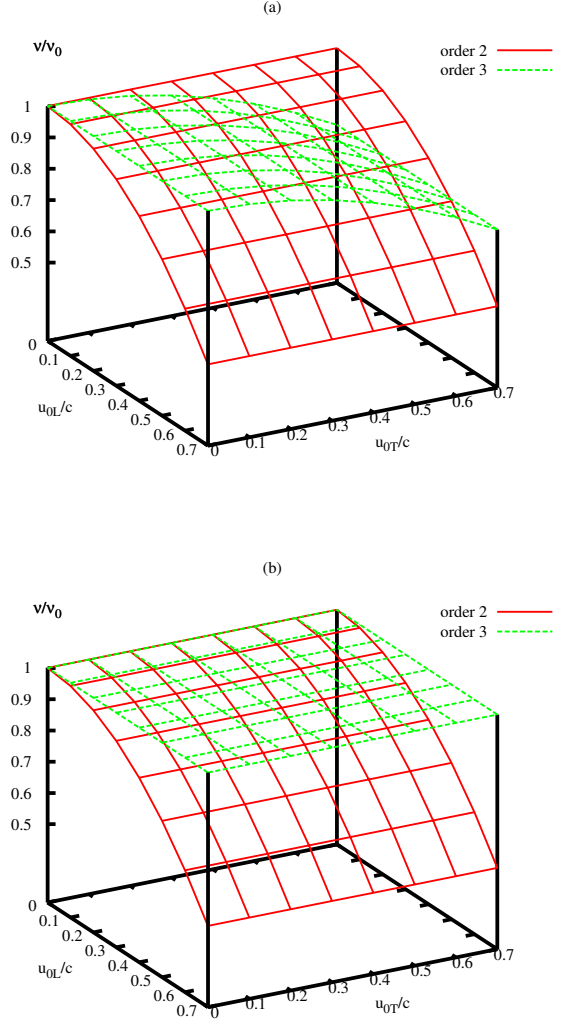}
  \caption{Velocity-dependence of viscosity of the fifth order D3Q19
    model (left) and the seventh order 39-velocity model (right) using
    second- and third-order equilibrium distribution function. Here,
    $\nu_0$ is the theoretical value of viscosity at zero mean
    velocity and $c = \sqrt{T_0}$ is the isothermal sound speed.}
  \label{fig1}
\end{figure}

Shown in Fig.~\ref{fig1} are the measured viscosity, normalized with
respect to its theoretical value, as a function of the longitudinal
and transverse components of $\bu_0$.  Here the magnitude of $\bu_0$
are expressed in terms of the corresponding Mach numbers.  On the top
are the results using the 19 point (D3Q19) model with second and third
order equilibrium distribution function ($N=2$ and $3$ respectively).
The dependency on the translational velocity is significant in both
cases except for small Mach numbers.  For the case of $N=2$, the error
is caused by the missing third order term in the equilibrium
distribution function.  For the case of $N=3$, the 19-velocity
quadrature is not sufficiently accurate for the third-order moments
and results in an error in another form.  On the bottom are results
using the 39-velocity model.  To be seen is that the velocity
dependence of the viscosity is removed once the third order term is
included in the equilibrium distribution function.

\begin{figure}
  \onefigure[viewport=50 50 554 928,width=240pt]{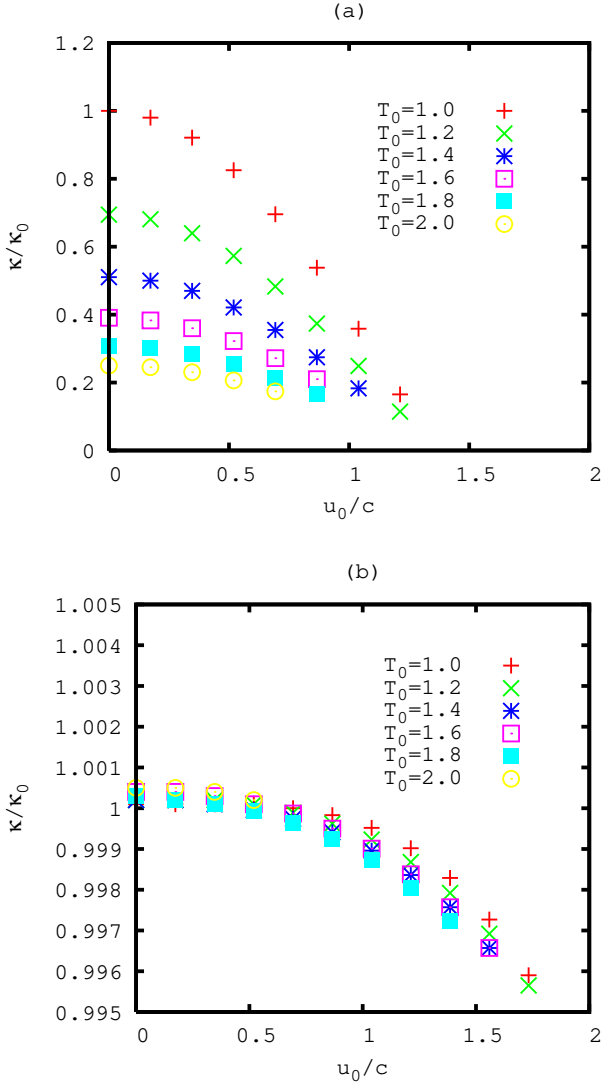}
  \caption{Velocity-dependence of thermal diffusivity for the 121
    point model.  The results in (a) and (b) are from the third-order
    and fourth-order equilibrium distribution function
    respectively. Here, $\kappa_0$ is the theoretical value of
    diffusivity at zero mean velocity.}
  \label{fig2}
\end{figure}

The thermal diffusivity is measured in a similar fashion.  Instead of
velocity, a initial sinusoidal temperature perturbation is imposed
while both the velocity and pressure are uniform.  Fig.~\ref{fig2}
shows the thermal diffusivity as functions of translational velocity
$u_0$ and the background temperature $T_0$.  Here $u_0$ is taken to
have the equal components in three directions.  As shown in
Fig.~\ref{fig2}a, with the third-order equilibrium distribution
function, the diffusivity varies with the translational
velocity. Furthermore, this dependence of velocity also depends on the
background temperature.  As shown in Fig.~\ref{fig2}b, when
fourth-order equilibrium distribution function is used, the thermal
diffusivity does not depend on the translational velocity anymore and
agrees well with the theoretical value for different back ground
temperatures.  The Galilean invariance is completely recovered.
However, the stability of the model depends one background temperature
and translational velocity.  Hence, in Fig.~\ref{fig2}, there are no
data for some high velocity cases. Detailed discussion of the
instability is beyond the scope of this paper.

\section{Discussion}

\label{sec:discussion}

In this Letter, we give the quantitative conditions for the LBGK
system to have the same macroscopic hydrodynamics as the continuum
kinetic system described by the BGK equation.  The degree of precision
of the velocity set as a Gauss-Hermite quadrature together with the
order of the Hermite terms retained in the distribution function
determines the order of the hydrodynamic moments that will have the
correct macroscopic behavior.  The velocity dependency of the
viscosity in the commonly used LBGK system is identified as an error
due to both the insufficient truncation of the distribution function
in moment space and the insufficient isotropy of the 9-velocity model.
Once sufficient moments are retained in the equilibrium distribution
and accurate quadrature is used, the hydrodynamic behavior of the LBGK
asymptotically approaches to that of the BGK equation in continuum.
Kinetic models for thermal fluids, fluid mixtures and fluids beyond
Navier-Stokes hydrodynamics can be constructed with guaranteed
Galilean invariance.

Centered in the effort of restoring full Galilean invariance in LBGK
system is the search for velocity set that makes isotropic tensors.
This effort is shown here to be equivalent to finding the sufficiently
accurate Hermite quadratures.  On a regular Cartesian grid, this task
was solved systematically~\cite{Shan06,Philippi06}.  To fully recover
the Galilean invariant Navier-Stokes momentum equation, sixth order
accurate quadratures are required.  It can be verified that with the
D2Q17 model~\cite{Qian98}, the diagonal component of the sixth tensor
is not isotropic, causing the Galilean invariance not fully restored
in the corresponding terms.  Through exhaustive search, it can be
concluded that to have the sixth order isotropy with velocities
forming a regular lattice, speed-3 velocities must be included. The
minimum velocity set found on a regular grid with sixth order isotropy
is the 39-velocity model in three dimensions and the 21-velocity model
in two dimensions. A different velocity set with sixth order isotropy has been given by Chikatamarla {\it et al}~\cite{Chikatamarla06}. However, smaller velocity sets not entirely on a regular grid are also available~\cite{Shan06}.

\acknowledgments

This work is supported in part by the National Science Foundation.

%\bibliography{lattice,misc}
%\bibliographystyle{eplbib}

\end{document}